\documentclass[%
 aip,
 amsmath,amssymb,
 preprint,%
]{revtex4-1}

\usepackage{graphicx}% Include figure files
\usepackage{dcolumn}% Align table columns on decimal point
\usepackage{bm}% bold math
\usepackage[utf8]{inputenc}
\usepackage[T1]{fontenc}
\usepackage{mathptmx}
\usepackage{etoolbox}
\usepackage{multirow}
\usepackage{xcolor}

\newcommand{\MSEED}{\langle R_{ee}^{2} \rangle}
\newcommand{\pknot}{P_\mathrm{knot}}

\makeatletter
\def\@email#1#2{%
 \endgroup
 \patchcmd{\titleblock@produce}
  {\frontmatter@RRAPformat}
  {\frontmatter@RRAPformat{\produce@RRAP{*#1\href{mailto:#2}{#2}}}\frontmatter@RRAPformat}
  {}{}
}%

\makeatother

\newcommand{\hmrev}[1]{\textcolor{black}{#1}}
\newcommand{\msrev}[1]{\textcolor{black}{#1}}

\begin{document}

\preprint{AIP/123-QED}

\title[]{Topological comparison of flexible and semiflexible chains in polymer melts with $\theta$-chains}
% Force line breaks with \\
\author{Maurice P. Schmitt}
\author{Sarah Wettermann}
 \affiliation{ 
Institut für Physik, Johannes Gutenberg-Universität Mainz, Staudinger Weg 9, 55099 Mainz, Germany
}
\author{Kostas Ch. Daoulas}
\affiliation{ 
Max Planck Institute for Polymer Research, Ackermannweg 10, 55128 Mainz, Germany
}
\author{Hendrik Meyer}
\affiliation{ 
Institut Charles Sadron, Université de Strasbourg, CNRS UPR 22, 23 rue du Loess-BP 84047, 67034 Strasbourg, France
}
\author{$^*$Peter Virnau}
 \affiliation{
Institut für Physik, Johannes Gutenberg-Universität Mainz, Staudinger Weg 9, 55099 Mainz, Germany
}
 \email{corresponding author virnau@uni-mainz.de \\ \ \  daoulas@mpip-mainz.mpg.de, hendrik.meyer@ics-cnrs.unistra.fr}

\date{\today}% It is always \today, today,

\begin{abstract}
A central paradigm of polymer physics states that chains in melts behave like random walks as intra- and interchain interactions effectively cancel each other out. Likewise, $\theta$-chains, i.e., chains at the transition from a swollen coil to a globular phase, are also thought to behave like ideal chains, as attractive forces are counterbalanced by repulsive entropic contributions. While the simple mapping to an equivalent Kuhn chain works rather well in most scenarios with corrections to scaling, random walks do not accurately capture the topology and knots particularly for flexible chains. In this paper, we demonstrate with Monte Carlo and molecular dynamics simulations that chains in polymer melts and $\theta$-chains not only agree on a structural level for a range of stiffnesses, but also topologically. They exhibit similar knotting probabilities and knot sizes, both of which are not captured by ideal chain representations.
\hmrev{This discrepancy comes from the suppression of small knots in real chains, which is strongest for very flexible chains because excluded volume effects are still active locally and become weaker with increasing semiflexibility.}
Our findings suggest that corrections to ideal behavior are indeed similar for the two scenarios of real chains and that structure and topology of a chain in a melt can be approximately reproduced by a corresponding $\theta$-chain.

\end{abstract}

\maketitle

\section{Introduction}
Polymers are macromolecules composed of repeating subunits with a wide range of properties and applications. Not only do synthetic polymers have an immense impact on the economy and our daily lives, but life itself is based on biopolymers, such as DNA, proteins and lipids \cite{Molecular_Biology_of_the_Cell}. While the interplay of the constituent atoms, which exhibit numerous interactions, appears complex and intractable at first glance, polymer physics \cite{Flory, deGennes,Doi,rubinstein:polymerphysics,WangReview2017macro} has developped successful tools to describe generic static and dynamic behavior of polymers in a coarse-grained manner. 

A fundamental principle of polymer physics is the idea that polymers can often be described as ideal chains in which non-bonded interactions between monomers are neglected \cite{Flory, deGennes,Doi,rubinstein:polymerphysics}. These descriptions not only work surprisingly well, but they are also attractive from a theoretical point of view as properties of ideal chains can generally be calculated analytically. Examples include long strands of unconfined DNA, which can be described to a good approximation by a worm-like chain model \cite{Kratky_1949, Marko_1995}. In this case, the chain extension is much larger than the segment diameter of the chain, and the large-scale behavior is governed by DNA rigidity rather than by monomer interactions on the local scale. Chains in polymer melts are also characterized by ideal chains. The concentration of monomers of a particular chain in the melt has a peak around its center of mass. A monomer of this chain thus experiences a repulsive potential which is proportional to the local concentration.\cite{deGennes} At the same time, the concentration of monomers of the rest of the melt exhibits a corresponding trough since the combined concentration profile is flat apart from fluctuations. Hence, repulsive forces acting on a monomer of a particular chain emerging from intrachain contacts are on average canceled out by attractive forces arising from interactions with neighboring chains. Thus, overall, a chain experiences no net forces and remains ideal.
A similar self-consistent field argument can be made for single chains at the $\theta$-transition point between a swollen state (representing good solvent conditions) and a globular phase (representing a bad solvent). Just at the transition, repulsive entropic forces cancel out with attractive forces, leading to ideal chain behavior again. Of course, theoretical arguments such as those outlined above need to be tested to reveal limitations and corrections to the ideal behavior. Fortunately, computer simulations provide the required tools.

Over the past 70 years, numerical simulations in general \cite{Allen, Frenkel_Smit} and Monte Carlo simulations \cite{Landau_Binder} in particular have emerged as a cornerstone in the ever-increasing building of physics strengthening and interacting with experiments and analytical theory alike. Shortly after the seminal publication of the Metropolis algorithm in 1953 \cite{metropolis1953}, the first computer simulations to grow self-avoiding polymer chains emerged \cite{Wall_1954, Rosenbluth_1955} and formed the basis for a class of chain-growth algorithms \cite{Laso_1992, Siepmann_1992, Grassberger_1997} still employed today. Over the years, polymers have become important target systems for a multitude of Monte Carlo algorithms, which also enable simulations of dense systems \cite{Laso_1992, Siepmann_1992, Mansfield:JCP:1982, Olaj:MMC:1982, Auhl2003EquilibrationOL, Reith:CPC:2010}. Related to our study, we would like to highlight yet another very successful algorithm to sample single chain conformations. Originally proposed by Lal in 1969 \cite{lal1969} and popularized in the late 1980s by Madras and Sokal \cite{madras1988}, the so-called pivot algorithm is based on the idea that global rotations of large polymer segments are able to decorrelate dilute chains within a few Monte Carlo steps. With optimized versions of this algorithm \cite{clisby2010}, simulations of single chains with lengths exceeding $10^7$ monomers are now feasible.

As simulations provide direct access to the microscopic structure, they are a powerful tool for testing assumptions and approximations entering analytical arguments.
In this study, we investigate with Monte Carlo and molecular dynamics simulations one of the central paradigms of polymer physics, namely that chains in melts and at the $\theta$-transition are well-described by ideal chains. Logarithmic corrections have been known for a long time but are supposed to play a role only for extremely long chains of several $10^4$ monomers\cite{HaSc99pre, Grassberger_1997}. 
Previous works have shown that there are systematic corrections to ideality in the melt where it was possible, thanks to MC methods, to sample chain lengths up to 8000.\cite{wittmer2004,wittmer2007,wittmer2011} These corrections show up as a power law in bond-bond correlation functions and are significant for the very flexible models often used in simulations (bond fluctuation and bead-spring model). Similar corrections have been found at the $\theta$-point \cite{ShPaLiRu2008macro,WangReview2017macro}. In the melt, these corrections emerge from the incompressibility\cite{beckrich2007} whereas in solution they are generated from a non-locality of interactions\cite{ShPaLiRu2008macro,ZhAlWa2020macro,schaefer1999book}. Both generate long-range correlations with a slower convergence to the ideal behavior, which is still supposed to hold asymptotically. 

Here, we attempt to contribute to this important topic from a topological point of view. 
Recent studies \cite{Rieger:PLoS:2016, Marenz_PRL2016, Zierenberg_Polymers_2016, Meyer:ACSMacro:2018:knots_melt, Wettermann_2023} indicate that the ability of a polymer chain to form knots \cite{Knots_in_soft_matter_review, grosberg2009few, Micheletti_PhRep2011, Dai:PRL:2015} can be considered as a fine gauge for its microscopic structure, comparable or even better than more traditional descriptors, since self-entanglements are very sensitive to the local and mesoscale shape of the polymer. Thus, we ask how well ideal chains reproduce the occurrence and sizes of knots in polymer melts and $\theta$-chains. 
While in previous work \cite{Meyer:ACSMacro:2018:knots_melt, Zhang_2020, Tubiana_2021} we have found an indication that ideal chain representations tend to overestimate the occurrence of knots in polymer melts, at least for flexible chains, here we extend our analysis to $\theta$-chains at various stiffnesses and a single-chain model that has the same size as a chain in a melt.
Intriguingly, while knots are not well-captured by ideal representations, chains in melts and $\theta$-chains exhibit comparable knotting behavior. This outcome suggests that similar corrections to the ideal behavior indeed apply to the two scenarios.

\section{Methods}
In this section, we provide an overview of our coarse-grained polymer model, simulation methods to obtain polymer melts and corresponding single chains, and tools used for structural and topological analysis.

\subsection{Reference polymer melts: model, molecular dynamics simulations and corresponding ideal chains}
Each coarse-grained polymer \cite{KremerGrest} consists of $N+1$ beads connected by $N$ harmonic springs with energy $V_\text{harm} = 400 \, k_B T/\sigma^2 (r_{i,i+1} - 0.967\sigma)^2$ to obtain an average bond length in line with the expectation value for the standard FENE potential.
Non-bonded monomers interact via a purely repulsive Weeks-Chandler-Andersen (WCA) potential, which corresponds to a Lennard-Jones potential truncated and shifted at the position of its minimum $r_{c}=\sqrt[6]{2}\sigma$. The chain stiffness is implemented with an angular potential of the form $V_B = B(1-\cos \theta_i)$, where $\theta_i$ is the angle between the bond vectors $\vec r_{i-1,i}$ and $\vec r_{i,i+1}$. In our study, $B$ ranges from 0 (fully flexible) to 4 (semi-flexible). 

Reference molecular dynamics simulations of polymer melts were obtained with the software package LAMMPS \cite{PLIMPTON19951} employing a standard Nose-Hover thermostat. 
Unless otherwise noted, a system with $768$ chains consisting of $1024$ monomers each at a monomer density of $\rho=0.68 \, \sigma^{-3}$ was simulated for each stiffness parameter $B$. To generate a start configuration, we generate random walks (RWs) of the freely rotating chain model (FRC) with an angle such that we obtain the desired characteristic ratio. These phantom chains are moved in space by MC steps to pre-equilibrate the monomer density. Then  the WCA potential is gradually added by increasing a force-cap parameter  \cite{Auhl2003EquilibrationOL}.  The system is further equilibrated by running for over five relaxation times, where a relaxation time is the time after which the end-to-end correlation is reduced to $10 \, \%$. The mean squared displacements behave as expected for reptating chains after this equilibration process.

Ideal chains representing chains in a melt are typically constructed by matching the contour length and mean square end-to-end distance of the underlying polymer model with those of a random walk \cite{rubinstein:polymerphysics}. To fulfill these two conditions, the two parameters $N^*$ and $l^*$ are adjusted, which represent the number of bonds (Kuhn segments) and bond length (Kuhn length) of the corresponding random walk, respectively. 
The mean squared end-to-end distance of a random walk is $N^* \times l^{*2}$, and the contour length is $N^* \times l^*$. Likewise, the contour length of the underlying model is $N \times \langle l \rangle$, and the mean square end-to-end distance $\langle R_\mathrm{ee}^2 \rangle$ is determined separately in simulations. Mapping these quantities yields a simple set of two linear equations, that can be solved for $l^* = \langle R_\mathrm{ee}^2 \rangle / (N \times \langle l \rangle) $ and $N^* = \langle R_\mathrm{ee}^2 \rangle / l^{*2}$. As such, corresponding random walks with the parameters $N^*$ and $l^*$ can be constructed for any polymer model with known $N$, $\langle l \rangle$ and $\langle R_\mathrm{ee}^2 \rangle$. 

\subsection{Monte Carlo simulations of $\theta$-chains}
To model transitions of single chains from a swollen coil to a globular phase \cite{Janke_2017}, we need to adjust the model described in the previous subsection to account for attraction. To this end, we extend the range of our non-bonded Lennard-Jones interaction by adjusting its cutoff to twice the distance at which the potential has its minimum ($2\sqrt[6]{2} \sigma$). For simplicity, we also fix the bond length to the minimum value of the harmonic bond potential (0.967). The transition to globular states can now be invoked by lowering the energy scale $\epsilon$ of the Lennard-Jones interaction to values less than unity, while keeping the temperature fixed.

Configurations were efficiently generated using a standard implementation of the pivot algorithm~\cite{lal1969, madras1988, clisby2010}. In each step, a monomer was chosen at random and one arm of the polymer was rotated by an arbitrary angle. After updating the conformation, the energy difference $\Delta E$ between the new and the old state was calculated and the conformation was accepted with the Metropolis criterion,~\cite{metropolis1953} i.e., if a drawn random number between zero and one was smaller than $e^{-\Delta E/k_B T}$. 

The finite-size estimates for the $\theta$-points of our single chain models at each stiffness $B$ were determined by running simulations at monomer numbers slightly above and below $N$.\cite{milchev1993off} We define the $\theta$-point (for a given $N$) as the point where the attractive and repulsive parts of real single chain models cancel out to yield the structural scaling of a random walk: $\langle R_\mathrm{ee}^2 \rangle \propto N$, where $ \langle R_\mathrm{ee}^2 \rangle$ is the mean squared end-to-end distance. Thus, the $\theta$-point can be identified by running simulations at two chain lengths above and below $N$ and finding the $\epsilon$ value for which the ratio $\langle R_\mathrm{ee}^2 \rangle / N$ becomes the same in both simulations (see supplementary material section \ref{appendix1} for more information.) For comparison, we also generated chains where $\langle R_\mathrm{ee}^2 \rangle$ is the same as in the melt, again, by varying $\epsilon$.

\subsection{Structural analysis and determination of knots}
The normalized single chain structure factor $P(q)$ and the normalized mean-square internal distance (nMSID) are used to analyze the structure of polymer chains obtained from simulations.

The structure factor is defined in a standard way as:
\begin{align*}
        P(q) &= \frac{1}{N^2}\Big\langle \Big| \sum_{i=1}^N \exp \left( {\rm i}\;\vec{q}\cdot \vec{R}_i\right) \Big|^2 \Big \rangle
\end{align*}

\noindent
where angular brackets denote an averaging over chain conformations and orientations of the scattering vector $\vec q $ (at fixed $|\vec q |$).

The nMSID is defined as $\mathrm{nMSID}(N_\Delta) = \langle R^2_{i,i+N_\Delta} \rangle_i / N_\Delta$, where $N_\Delta$ is the number of particles between which distances are considered, $R^2_{i,i+N_\Delta}$ is the squared distance between particle $i$ and $i+N_\Delta$, and $\langle ... \rangle_i$ implies an average over all $i$ such that $i+N_\Delta \le N$ and over all configurations accumulated during the simulation.

\begin{figure*}[ht!]
\centering
\includegraphics[width=0.5\textwidth]{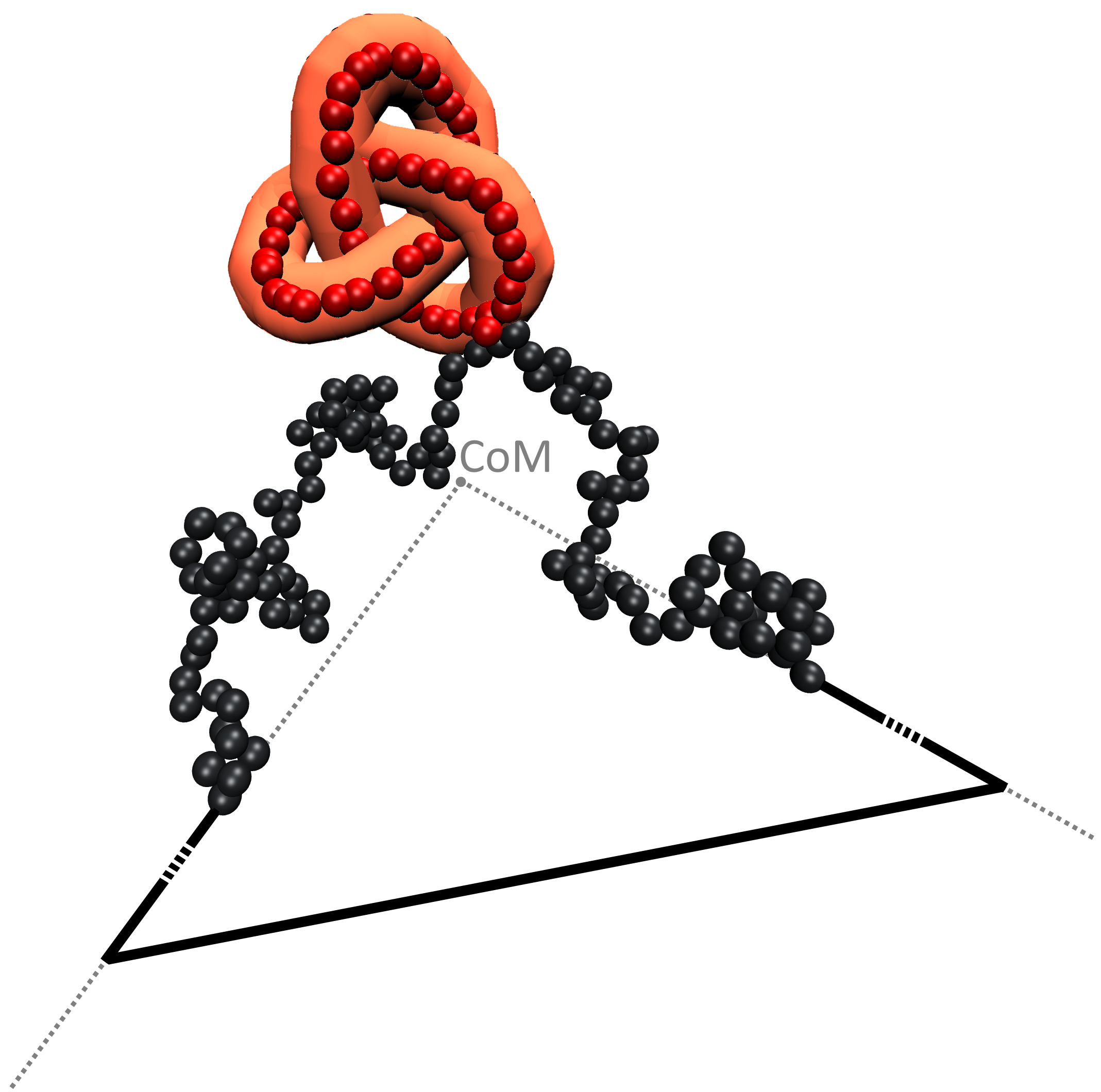}
\caption{Schematic drawing of the center-of-mass closure. The artificial closure (solid black lines, dashes imply a quasi-infinite extension) is drawn from the center of mass (CoM) through both termini and connected far away from any chain segments. A trefoil knot (red, enclosed in orange tube) is detected, while the black monomers do not contribute to the topology of the chain. For a comparison of closure schemes see e.g. Ref.~\cite{Tubiana:PTPS:2011}.}
\label{closure-sketch}
\end{figure*}

Knots are mathematically defined in closed loops \cite{Adams:1994}. Here, we draw two lines from the center-of-mass of the polymer in question to the respective termini \cite{Mansfield_1997, virnau2006, virnau2010}, which define the direction of our connecting lines. The latter emerge from the termini and continue outward before they are connected far away from the polymer as shown schematically in Fig.~\ref{closure-sketch}. Once connected, we compute a variant of the Alexander polynomial~\cite{virnau_2010}, which suffices to distinguish between simple knots such as the unknot and the trefoil ($3_1$) knot (displayed in Fig.~\ref{closure-sketch}) considered in this study. Knot sizes are determined by successively removing beads from either end before closure until the knot type changes~\cite{Kantor}.

\section{Results}
\subsection{Structural properties}
\begin{figure*}[ht!]
\centering
\includegraphics[width=0.9\textwidth]{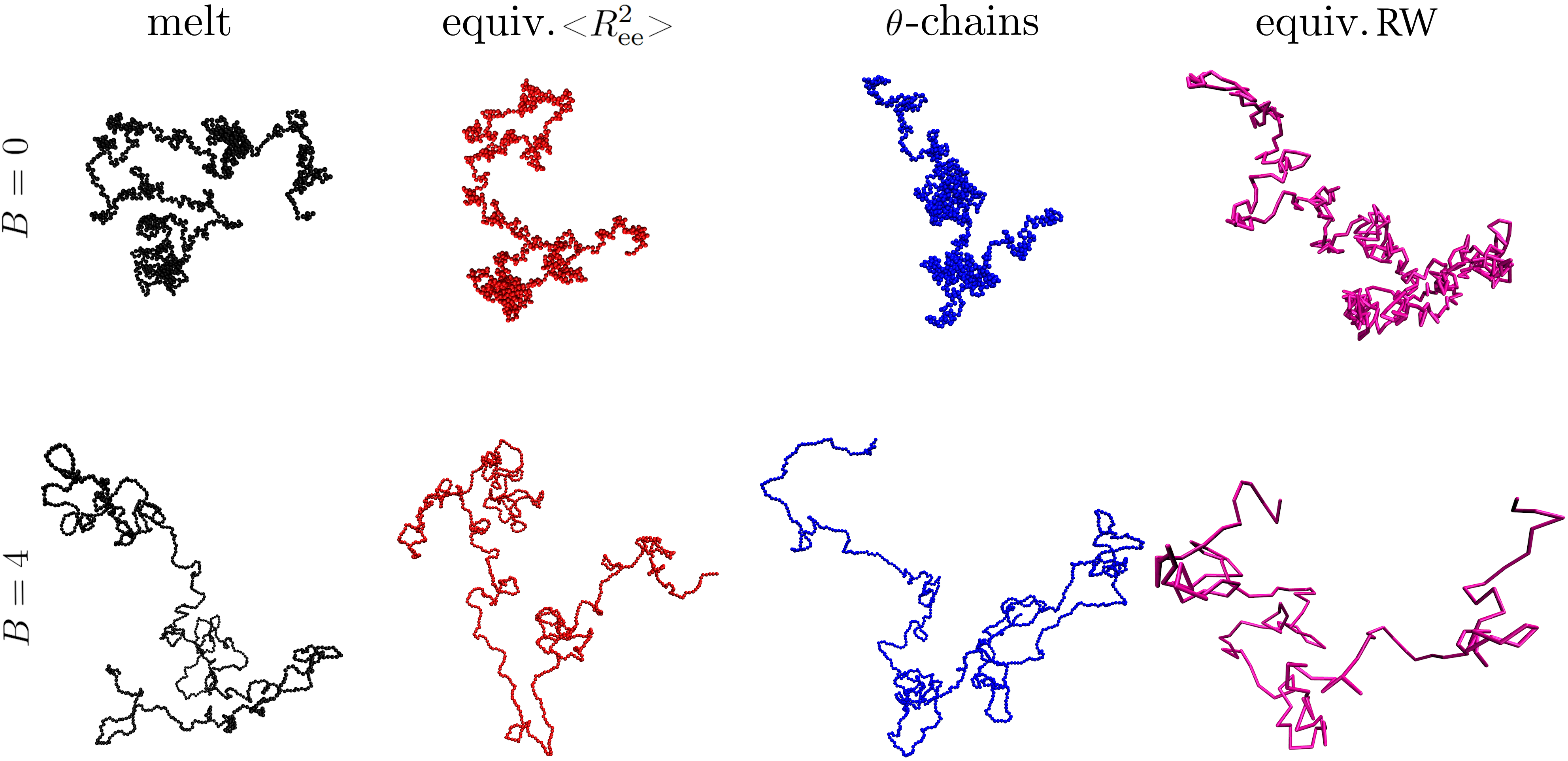}
\caption{Comparison of snapshots of melts and single chain models without ($B=0$) and with ($B=4$) a bending potential. The melt chains and real chains consist of $1024$ monomers, while the corresponding monomer number of the random walks is stiffness-dependent. The configurations shown have been chosen so that their gyration radii are similar to the mean gyration radii. The conformations of chains in the melt show strong similarities to the single chain models when the bending constant $B$ is equal. The equivalent random walks are represented by cylinders between steps and consist of fewer monomers, but exhibit qualitative similarities on large scales.}
\label{trajectory_collection}
\end{figure*}
\noindent
It is often assumed that polymers in melts form random walks as hypothesized by Flory \cite{Flory}. Indeed, snapshots of chain configurations from a melt compared to single chain models show considerable qualitative similarities between the melt and other single chain models such as the random walk, as shown in Fig. \ref{trajectory_collection}. The conformations at $B=0$ form more clusters on a small scale due to their flexibility, whereas stiffer $B=4$ chains tend to form straight rods on small scales; therefore, the equivalent random walk at $B=4$ exhibits a significantly larger bond length and consists of significantly fewer beads.

\begin{figure*}[ht!]
\centering
\includegraphics[width=\textwidth]{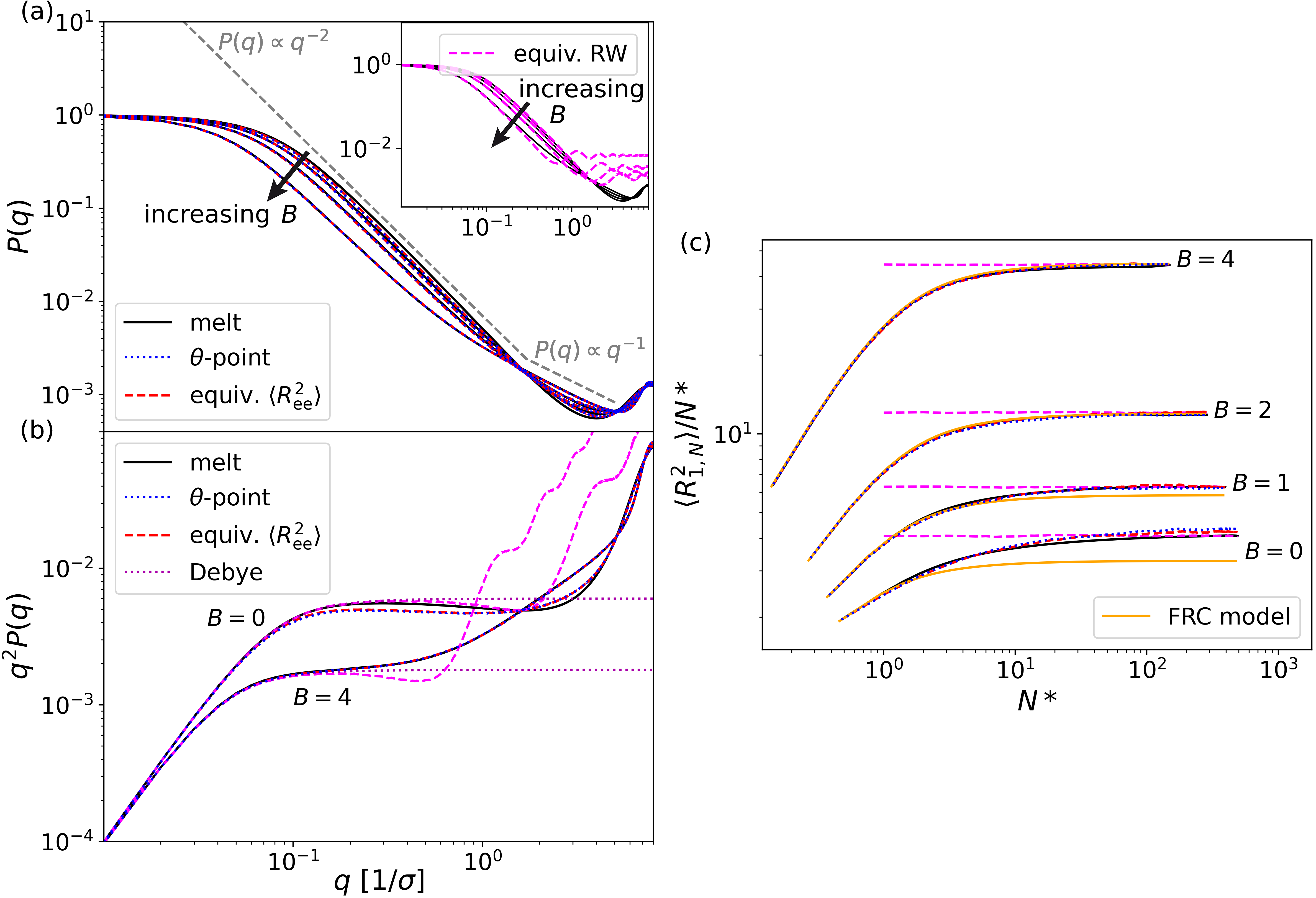}
\caption{Structure of chains in polymer melts compared to single chain models. The melt and real chains consist of $1024$ beads, although their length is converted to the number of Kuhn segments $N^*$. The stiffness is varied from $B=0, 1, 2, 4$. (a) Single chain structure factors $P(q)$ \textcolor{black}{for melt chains (black lines), $\theta$-chains (dotted blue lines) and equivalent $\langle R_{\mathrm{ee}}^2 \rangle$-chain (dashed red lines) for $B=0,1,2,4$}. All $P(q)$ cross at $q \approx 1.7 \, \sigma^{-1}$, where the resolved structure of the stiffest $B=4$ chains has a crossover from ideal chain behavior $P(q) \propto q^{-2}$ to rod-like behavior $P(q) \propto q^{-1}$, as indicated by the gray dashed line as a guide to the eye. The inset shows a separate comparison of the melt to the equivalent random walks \textcolor{black}{denoted by pink dashed lines. In both cases arrows indicate the direction of increasing $B$}. (b) Kratky representation, $q^2 P(q)$ vs. $q$, omitting $B=1$ and $B=2$ for clarity. The Debye function (purple dotted line) shows the analytically expected behavior for ideal chains up to $q \approx 2 \pi /l^*$. (c) Normalized mean squared internal distance (nMSID) $\langle R_{1,N}^2 \rangle /N^*$. The freely rotated chain (FRC) model (orange) shows better agreement with other models with increasing stiffness $B$.}
\label{Fig:structural_properties}
\end{figure*}

In order to examine the validity of Flory's hypothesis \cite{Flory} more closely, however, quantitative structural analyses are required. The single chain structure factor $P(q)$ of each model is shown in Fig. \ref{Fig:structural_properties}(a). From the initial plateau it decays with a power law $q^{-2}$ as expected from the Debye function for random walks. The chains with higher local rigidity also have a regime of $q^{-1}$ on scales of the Kuhn segment before increasing to the monomer peak which is not shown in the figure.
The figure reveals that the melt is structurally similar to all considered real single chain models at all wave numbers $q$, i.e., $P(q)$ shows structural agreement at all length scales, implying that a mapping of the melt to the single chain models works well on this scale.

However, the random walk shown in the inset of Fig.~\ref{Fig:structural_properties}(a) captures only the regime of $q^{-2}$ since the internal structure of a Kuhn segment is absent.
Fig.~\ref{Fig:structural_properties}(b) shows the single chain structure factor in Kratky representation where a random walk shows the so-called Kratky-plateau, a horizontal line. The figure contains the Debye function which describes the structure factor of an ideal continuous random walk where the radius of gyration 
matches the radius of gyration of polymers in the simulations of melts. This representation makes the corrections to ideality \cite{wittmer2004,wittmer2011,beckrich2007} visible for flexible chains where the structure factor decays with respect to the Kratky plateau. For stiffer chains, this effect is so weak that it is hidden by the increase from the local persistence. The magenta dashed line is of our simulated random walk. Its fixed constant bond length induces oscillations for large $q$.

The nMSID in Fig. \ref{Fig:structural_properties}(c) supports the discussed findings of $P(q)$ and presents them in non-reciprocal space: The equivalent Kuhn RW is now just a horizontal line. All real chain models converge to this plateau by construction --- at sufficiently large $N$ all models do indeed scale like random walks, $\langle R_\mathrm{ee}^2 \rangle \propto N$. This is consistent with the $P(q) \propto q^{-2}$ regime observed at smaller $q$ in panel (a). 
At smaller scales, i.e. small $N^* \leftrightarrow$ large $q$, there are significant discrepancies between the random walk and the other models, i.e. melt and single chain models, due to the discussed influence of self-avoiding interactions as well as differing monomer number. We have plotted also the freely rotating chain (FRC) model which is still an ideal RW (phantom chain) that, however, takes into account local angular rigidity\cite{rubinstein:polymerphysics}. The local angular rigidity leads to a Flory characteristic ratio that converges to a constant value with inreasing the chain length as $N^{-1}$. We choose the average angle between two consecutive bonds as a parameter of the FRC expression. For large $B$, taking into account the local bending rigidity seems to be sufficient for matching the entire curve of the real chains, meaning that excluded volume effects are negligible.
For $B=0$ and $B=1$ there is a significant difference between the plateau of the FRC curve and the equivalent Kuhn chain. This difference comes from non-local correlations with bonds further away which recursively make the corrections to ideality converging more slowly as $N^{-0.5}$ (see refs.~\cite{wittmer2004,wittmer2007,wittmer2007b,wittmer2011,ShPaLiRu2008macro,WangReview2017macro}).

\subsection{Topological properties}
\begin{figure}[ht!]
\centering
\includegraphics[width=0.55\textwidth]{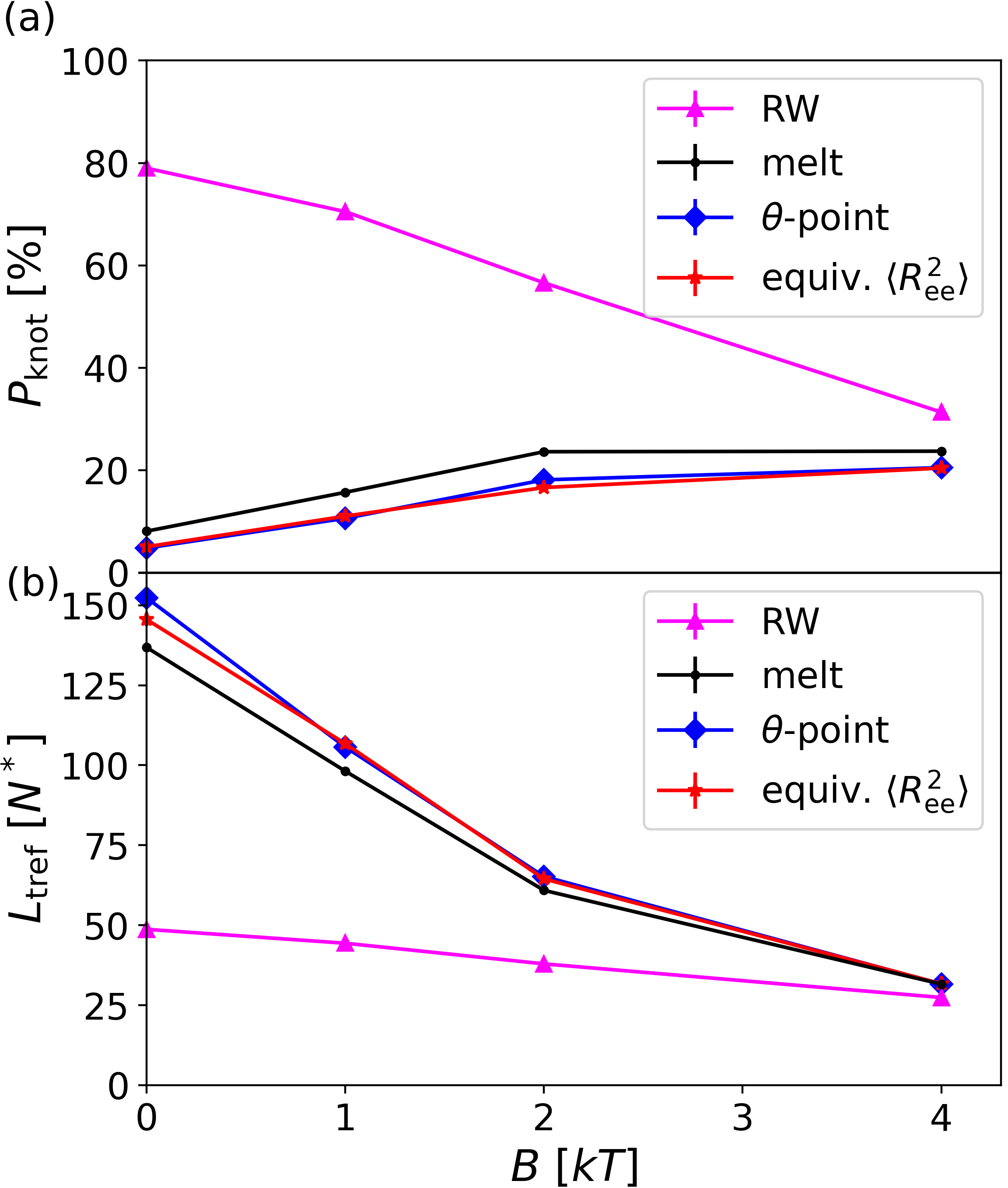}
\caption{Topological properties of polymer melts compared to single chain models. For these real-chain models, (a) the knotting probability $\pknot$ increases with increasing stiffness, while (b) the average trefoil knot length shrinks. Note that the melt and equivalent $\theta$-point chains have all the same number of monomers, whereas the random walks become shorter with increasing $B$. $\pknot$ for the equivalent ramdom walk thus decreases. \msrev{The solid lines connecting data points are a guide to the eye.}}
\label{fig:topology_vs_B}
\end{figure}

Figure~\ref{fig:topology_vs_B} reveals the main result of our study. In Fig.~\ref{fig:topology_vs_B}a we present knotting probabilities of chains in a polymer melt as a function of chain stiffness and compare them to a corresponding random walk, a chain at the (finite-size) $\theta$-point, and a single chain with the same $\langle R_\mathrm{ee}^2 \rangle$ as a chain in the melt. As already noted in Ref.~\cite{Meyer:ACSMacro:2018:knots_melt} random walks vastly overestimate the occurrence of knots in melts, particularly for flexible chains. Since random walks exhibit numerous localized knots, even remnants of self-avoidance suffice to suppress knotting at local scales. 
However, with increasing stiffness, the difference between real and ideal chain becomes weaker. For $B=4$ the values are actually rather close to each other.

The real chains at the $\theta$-point and from the melt follow the same trend and are rather close to each other. However, there is a notable difference:
The knotting probabilities of real chains exhibit better relative agreement with the melt in the semiflexible $B=4$ case $(P_\mathrm{knot}^\mathrm{melt} - P_\mathrm{knot}^\theta)/P_\mathrm{knot}^\mathrm{melt} \approx 13.6 \%$ than in the flexible $B=0$ case $(P_\mathrm{knot}^\mathrm{melt} - P_\mathrm{knot}^\theta)/P_\mathrm{knot}^\mathrm{melt} \approx 38.8 \%$. The same behavior is mirrored when sizes of trefoil knots are considered (Fig.~\ref{fig:topology_vs_B}b). The improvement of the agreement between knotting properties in single chain models and reference melts as the polymers become stiffer is consistent with an earlier study of knotting properties in polymer melts described with a mesoscopic (soft) model.~\cite{Zhang_2020} While the sizes of our two single chain models almost agree quantitatively with the ones measured for melt chains,  knots in the equivalent random walks are much smaller on average.

\begin{figure}[ht!]
\centering
\includegraphics[width=0.55\textwidth]{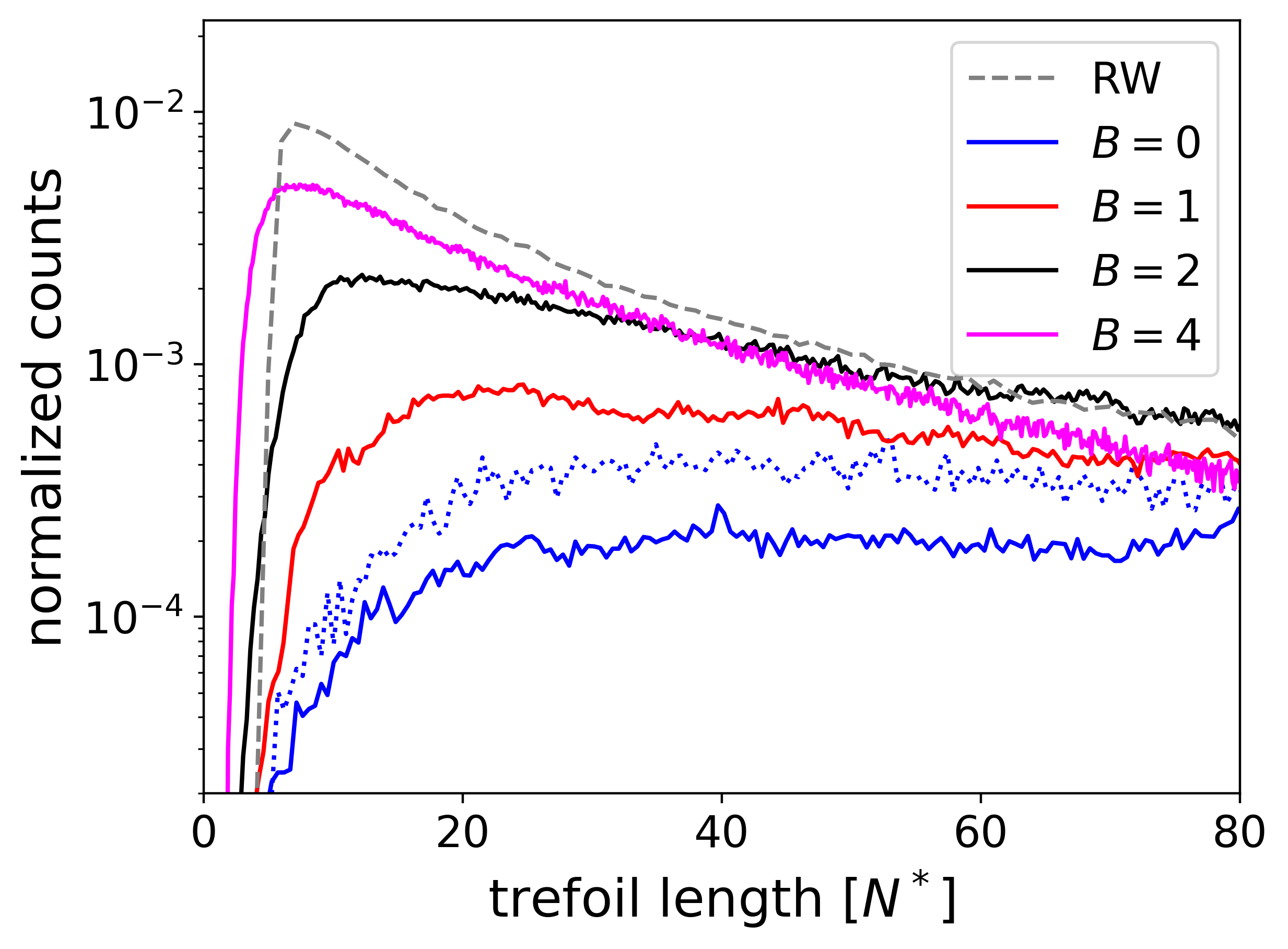}
\caption{Distribution of trefoil sizes in $\theta$-chains (solid lines) at varying stiffness $B$, of the $B=0$ melt (blue dotted line) and the RW trefoil size distribution (gray dashed line). The normalization was chosen such that the integral over the probability distribution yields the probability of obtaining a trefoil knot. In the case of the random walk, we used a size corresponding to $B=0$. Random walks form more tight knots spanning less Kuhn segments $N^*$ than melts or $\theta$-chains. The real chain distributions exhibit better agreement with the RW when stronger rigidity is introduced.}
\label{fig:treflen_distribution}
\end{figure}
One can understand this trend when looking at the distribution of the trefoil knot size shown in Fig.~\ref{fig:treflen_distribution}. Random walks have only a weak chain length dependence, we plot the result of the longest only as a gray dashed line. It has a maximum below ten Kuhn segments. This means that these small trefoil knots are composed of almost straight lines. The distributions of $\theta$-chains are shown as continuous lines. The trefoil knot size distribution of the stiffest $\theta$-chain $B=4$ is still quite close to the RW, but with decreasing $B$, small knots (in terms of Kuhn lengths) are increasingly avoided. The same holds for melt chains (for clarity, the figure shows the melt data only for $B=0$, for the higher rigidities, they are even closer to the $\theta$-chain distribution). This can be understood with the local swelling of flexible chains: The flexible chain is "bulkier," and the Kuhn length is comparable to the monomer diameter. This means in the pervaded volume of the Kuhn segment of a flexible polymer, there is not much space for the chain to come back to itself to form a knot, which suppresses the probability of finding knots consisting of few Kuhn segments (and suppresses the overall knotting probability).  %The flexible chain is "bulkier", the ratio of Kuhn length and monomer diameter is small. This means in the pervaded volume of the Kuhn segment of a flexible polymer, there is merely space for one or two other segments whereas for a semiflexible chain there could be many other segments. This makes it also easier to come back on itself to form a knot (on the scale above the Kuhn segment). Note that the argument describes the same origin which leads to the corrections to ideality. However, it is not the long-range nature of correlations which matters for the suppression of knots, but the local avoidance which removes a large amount of knots occurring in phantom chains.

\begin{figure}[ht!]
\centering
\includegraphics[width=0.55\textwidth]{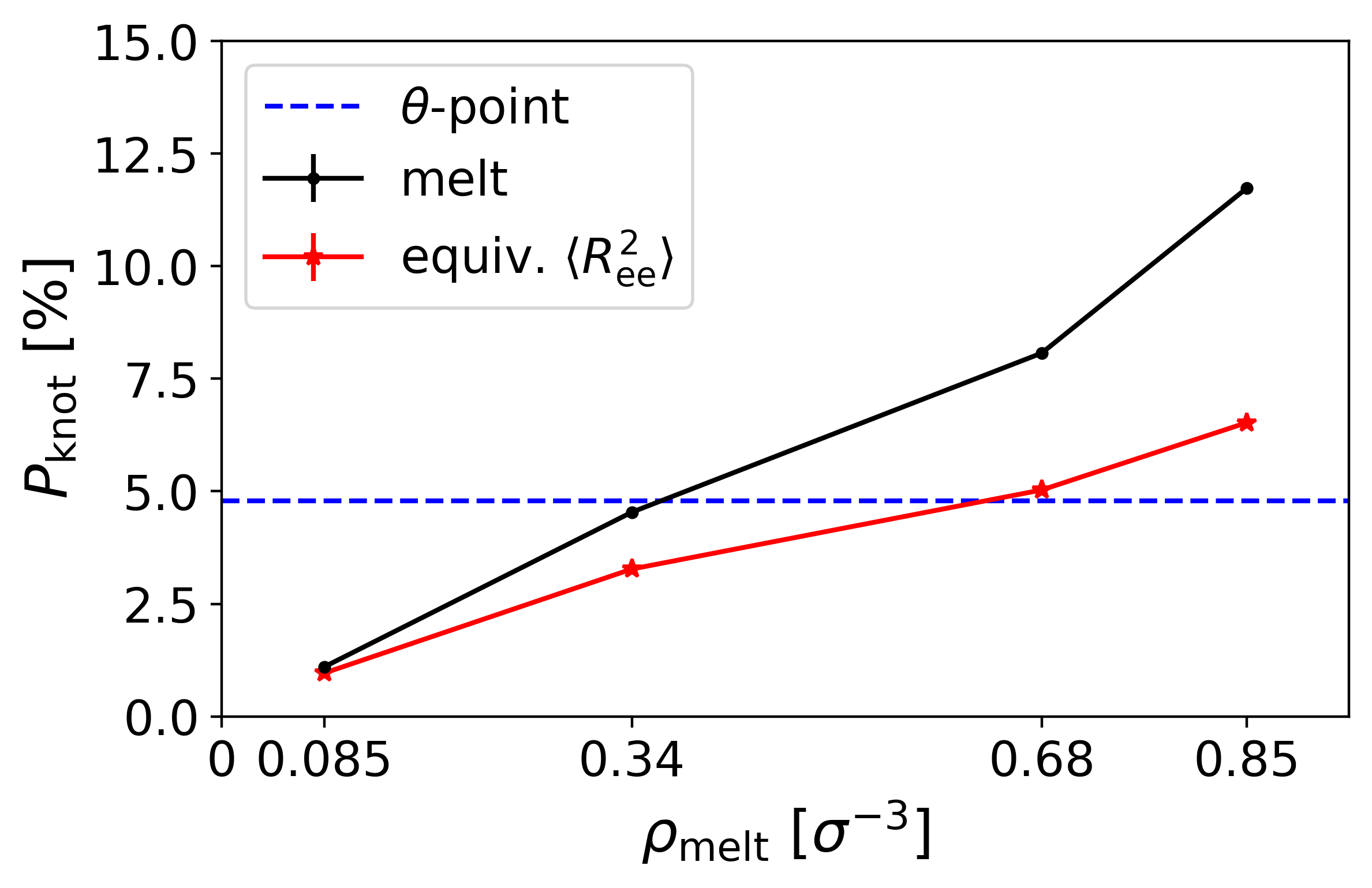}
\caption{Knotting probabilities of polymer melts at various densities and stiffness $B=0$ compared to single chain models with equivalent $\langle R_\mathrm{ee}^2 \rangle$ and the corresponding $\theta$-chain model. Single chains with equivalent $\langle R_\mathrm{ee}^2 \rangle$ show qualitative agreement with the melt, while the $\theta$-point is unique and thus cannot be varied to match density dependent melt properties. \msrev{The lines connecting data points act as a guide to the eye.} }
\label{fig:topology_vs_rho}
\end{figure}
Instead of ideal (phantom) chains, one should thus use real ($\theta$) chains to map melt conformations. Fig.~\ref{fig:topology_vs_rho}, however, explores the limitation of this mapping. Here, we compare the knotting probabilities of flexible melt chains ($B=0$) with $\theta$-chains and chains with the same $\langle R_\mathrm{ee}^2 \rangle$ as a function of density. As expected we observe almost perfect agreement for the latter at low densities, which worsens progressively with increasing melt density. At the highest density ($\rho=0.85$), the knotting probability for chains in the melt is almost twice the probability observed for single chains with the same $\langle R_\mathrm{ee}^2 \rangle$. Overall, the best agreement between $\theta$-chains and chains in the melt is achieved at densities $\rho\approx0.35$ in this particular scenario.
This can be understood as follows: Although melt and $\theta$-chains have the same leading order correction to the finite chain length characteristic ratio \cite{wittmer2004,ShPaLiRu2008macro}, the physical origin and thus the prefactors are not the same. This is related to the incompressibility for the melt and to non-local correlations at the $\theta$-point.    

\section{Conclusions}
We have used structural and topological quantifiers to investigate how polymer chains in a melt and chains at the $\theta$-transition point differ from ideal chain representations. These discrepancies are most severe for flexible chains, as remnants of self-avoidance dominate small-scale structures of polymer melt conformations and $\theta$-chains. While tight knots appear regularly in corresponding random walks, they are rare and only weakly localized in the two real chain scenarios. At these local scales, the chains also differ structurally from their idealized representations, as indicated, e.g., by the internal distance plot. The topological discrepancies become less pronounced as the chain stiffness increases and the local structure becomes dominated by stiffness rather than self-avoidance. Interestingly, $\theta$-chains and chains in a melt not only exhibit very similar local structure over all stiffnesses considered in this study, but also similar knotting probabilities and sizes indicating similar corrections to ideal behavior. 

Our results also suggest that, within limits, the structural behavior of chains in a melt can indeed be probed approximately by simulating single chains with the same end-to-end distance. These conformations are readily available and (unlike polymer melts) 
can be probed inexpensively using advanced Monte Carlo algorithms. 
\textcolor{black}{Slip-link models (SLM)~\cite{HuaSL,MasubuchiSL,LikhtmanSL,ChappaSL,SchneiderSL,MegariotisSL} for rheological studies are an example where our findings might be directly applicable. Especially, single-chain SLM~\cite{HuaSL,LikhtmanSL} are interesting as computational analogies of the tube model.~\cite{Doi} 
In these cases, a single polymer chain is "tethered" to the background by transient elastic springs which do not affect the equilibrium conformational statistics but constrain the motion of the chain during its dynamical evolution.~\cite{LikhtmanSL} 
Using SLM $\theta$-chains instead of ideal chains~\cite{LikhtmanSL} should enable these models to explicitly reproduce the correct statistics of self-entanglements in melts and explore their effect on polymer dynamics and rheology. However, we anticipate that this probably has only minor influence (at least in quiescent conditions) and that knots form and unform via the usual reptation dynamics.}

\section{Supplementary Material}
\textcolor{black}{See the supplementary material for details on the method to determine the $\theta$-point and tables listing simulation parameters and quantitative analysis results.}

\begin{acknowledgments}
The authors gratefully acknowledge the computing time granted on the supercomputer MOGON II and III at Johannes Gutenberg University Mainz as part of NHR South-West. M. P. S. and S. W. are grateful to the Deutsche Forschungsgemeinschaft (DFG) for funding (SFB TRR 146, project number \#233630050). HM acknowledges a generous grant of cpu time on the HPC cluster CAIUS of the mesocenter of University of Strasbourg as well as discussions with A. Johner and J. Wittmer.
\end{acknowledgments}

\section*{AUTHOR DECLARATIONS}
\subsection*{Conflict of Interest}
The authors have no conflicts to disclose.

\section*{Data Availability Statement}
The data that support the findings of this study are available from the corresponding author upon reasonable request.

\section*{Credit Line}
This article may be downloaded for personal use only. Any other use requires prior permission of the author and AIP Publishing. This article appeared in \cite{schmitt2024topological} and may be found at https://pubs.aip.org/aip/jcp/article/161/14/144904/3316214/Topological-comparison-of-flexible-and.

\bibliography{main.bib}

\clearpage
\section{Supplementary Material}
\setcounter{page}{1}
\subsection{Determination of the $\theta$-point}
\label{appendix1}
\begin{figure*}[ht!]
\centering
\includegraphics[width=0.9\textwidth]{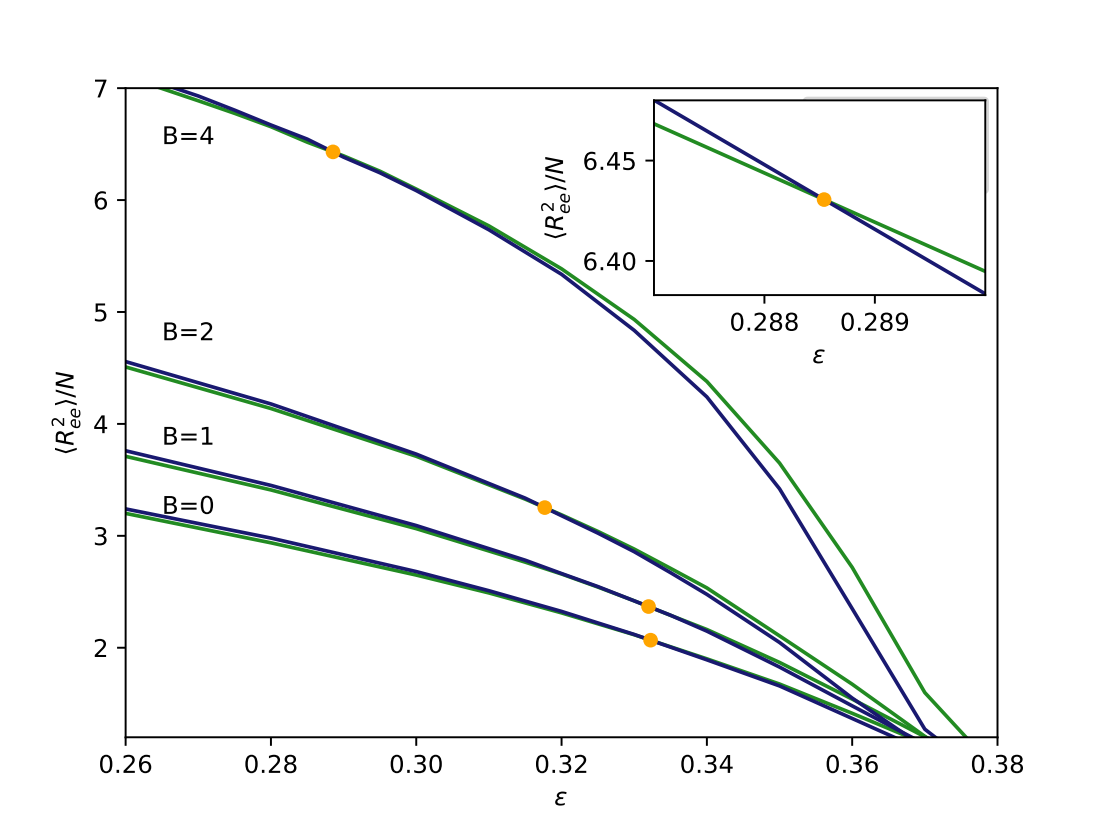}
\caption{Determination of the $\theta$-point. Two real chain simulations are performed at chain lengths slightly above and below $N+1=1024$ for various $\epsilon$. The point at which the normalized mean squared end-to-end distance of both simulations cross is the $\theta$-point, since the definition of the $\theta$-point, i.e. $\MSEED \propto N$, is fulfilled for the target chain length at exactly this point. The inset zooms in on the intersection for $B=4$.}
\label{theta-determination}
\end{figure*}
The $\theta$-point is defined by the point at which the mean squared end-to-end distance $\MSEED$ scales like an ideal chain, i.e. when $\MSEED$ is linearly proportional $\MSEED \propto N$ to the chain length.
In a real chain model with partially attractive pair-interactions, this is achieved by adjusting the ratio of the thermal energy $k_BT$ to the depth of the attractive potential, in our case $\epsilon$ from the Lennard-Jones potential; strong attractive pair-potentials relative to the thermal energy lead to globular states $R_g \propto N^{1/3}$, whereas high temperatures relative to the attractive pair-potentials lead to more extended configurations, where the interplay between entropy and self-avoiding interactions results in $R_g \propto N^{3/5}$.
At the $\theta$-point, the thermal energy and the attractive pair-potentials are in balance such that the scaling $\MSEED \propto N$ corresponds to that of ideal chains.

\clearpage
\subsection{Simulation parameters and results}
\label{appendix2}

\begin{table*}[ht!]
\centering
\begin{tabular}{ |c|c|c|c|c|c|c| } 
\hline
 & $B$ & $N^*$ & $\epsilon$ & $\langle R_\text{ee}^2 \rangle / N$ & $P_{\text{knot}}$ [\%]  & $L_\mathrm{tref}$ [$N^*$]\\
\hline
% MELT SECTION OF THE TABLE:
\multirow{4}{5em}{Polymer Melt} & 0 & 489 & -- & 1.953 & 8.07 & 136.8 $\pm$ 1.5\\ 
 & 1 & 394 & -- & 2.425 & 15.66 & 98.2 $\pm$ 0.7  \\ 
 & 2 & 284 & -- & 3.359 & 23.6 & 60.9 $\pm$ 0.4\\ 
 & 4 & 148 & -- & 6.455 & 23.7 & 31.6 $\pm$ 0.2 \\ 
\hline
% THETA SECTION OF THE TABLE:
\multirow{4}{5em}{$\theta$-Point}
 & $0$  & --  & $0.3322$  & $2.060 \pm 0.004$  & $4.782 \pm 0.012$  & $152.3 \pm 1.5$  \\
 & $1$  & --  & $0.3320$  & $2.369 \pm 0.005$  & $10.586 \pm 0.018$  & $105.7 \pm 1.0$  \\
 & $2$  & --  & $0.3177$  & $3.254 \pm 0.005$  & $18.099 \pm 0.025$  & $65.2 \pm 0.4$  \\
 & $4$  & --  & $0.2885$  & $6.453 \pm 0.004$  & $20.495 \pm 0.025$  & $31.6 \pm 0.1$  \\
\hline
% REQUIV SECTION OF THE TABLE:
\multirow{4}{5em}{$\langle R_\text{ee}^2 \rangle$ equiv. to the melt} 
 & $0$  & --  & $0.3377$  & $1.949 \pm 0.006$  & $5.030 \pm 0.018$  & $145.6 \pm 2.1$  \\
 & $1$  & --  & $0.3306$  & $2.388 \pm 0.007$  & $11.007 \pm 0.026$  & $106.7 \pm 1.5$  \\
 & $2$  & --  & $0.3135$  & $3.373 \pm 0.006$  & $16.600 \pm 0.032$  & $64.5 \pm 0.6$  \\
 & $4$  & --  & $0.2883$  & $6.452 \pm 0.005$  & $20.400 \pm 0.036$  & $31.7 \pm 0.1$  \\
\hline
% RW SECTION OF THE TABLE:
\multirow{4}{5em}{Random Walk}
 & $0$  & $489$  & --  & $1.954 \pm 0.006$  & $78.99 \pm 0.07$  & $48.7 \pm 0.1$  \\
 & $1$  & $394$  & --  & $2.430 \pm 0.009$  & $70.49 \pm 0.19$  & $44.4 \pm 0.2$  \\
 & $2$  & $284$  & --  & $3.364 \pm 0.006$  & $56.59 \pm 0.15$  & $37.9 \pm 0.2$  \\
 & $4$  & $148$  & --  & $6.459 \pm 0.008$  & $31.32 \pm 0.08$  & $27.4 \pm 0.1$  \\
\hline

\end{tabular}
\caption{Simulation parameters used and the corresponding results, as depicted in Fig. \ref{Fig:structural_properties} and Fig. \ref{fig:topology_vs_B}, i.e. for $N=1023$ and $\rho = 0.68 \, \sigma^{-3}$.}
\label{tab:theta-determination}
\end{table*}

\clearpage
\begin{table*}[ht!]
\centering
\begin{tabular}{ |c|c|c|c|c|c| } 
\hline
 & $\rho$ & $\epsilon$ & $\langle R_\text{ee}^2 \rangle / N$ & $P_{\text{knot}}$ [\%] \\
\hline
% MELT SECTION OF THE TABLE:
\multirow{4}{5em}{Polymer Melt}
& 0.085 & -- & 3.610 & 1.10 \\ 
& 0.34 & -- & 2.430 & 4.54  \\ 
 & 0.68 & -- & 1.953 & 8.07  \\ 
 & 0.85 & -- & 1.742 & 11.73 \\ 
 \hline
 % THETA SECTION OF THE TABLE:
 \multirow{1}{5em}{$\theta$-Point} & --  & $0.3322$  & $2.060 \pm 0.004$  & $4.782 \pm 0.012$ \\
 \hline
 % REQUIV SECTION OF THE TABLE:
\multirow{4}{5em}{$\langle R_\text{ee}^2 \rangle$ equiv. to the melt} 
& 0.085 & 0.22488 & $3.589 \pm 0.002$ & $0.963 \pm 0.008$\\ 
& 0.34 & 0.31327 & $2.430 \pm 0.004$ & $3.273 \pm 0.015$\\ 
& 0.68  & 0.3377 & $1.949 \pm 0.006$  & $5.030 \pm 0.018$ \\
& 0.85 & 0.34649 & $1.733 \pm 0.009$ & $6.524 \pm 0.021$\\ 
\hline

\end{tabular}
\caption{Simulation parameters used and the corresponding results, as depicted in Fig. \ref{fig:topology_vs_rho}, i.e. for $N=1023$ and $B=0$. There is only one $\theta$-point for single chains with the given $N$ and $B$; however, $\langle R_\mathrm{ee}^2 \rangle$ can be matched for all melt densities.}
\label{tab:rho-dependence}
\end{table*}

\end{document}